\documentclass[fleqn,usenatbib]{mnras}

\usepackage{newtxtext,newtxmath}
\usepackage[T1]{fontenc}
\usepackage{ae,aecompl}
\usepackage{booktabs,caption}
\usepackage[flushleft]{threeparttable}

\usepackage{graphicx}	% Including figure files
\usepackage{amsmath}	% Advanced maths commands
\usepackage{amssymb}	% Extra maths symbols
\usepackage{array,booktabs}
\usepackage{siunitx}
\title[M15 and M30 mass function]{Radial variation of the stellar mass functions in the  globular clusters M15 and M30: clues of a non-standard IMF?}

\author[M. Cadelano et al.]{
    M. Cadelano$^{1,2}$\thanks{E-mail: mario.cadelano@unibo.it}, 
    E. Dalessandro$^{2}$, 
    J. J. Webb$^{3}$, 
    E. Vesperini$^{4}$, 
    D. Lattanzio$^{1}$,
    G. Beccari$^{5}$, 
    \newauthor
    M. Gomez$^{6}$ 
     and L. Monaco$^{6}$
\\
$^{1}$Dipartimento di Fisica e Astronomia, Via Gobetti 93/2 I-40129 Bologna, Italy\\
$^{2}$INAF-Astrophysics and Space Science Observatory Bologna, Via Gobetti 93/3 I-40129 Bologna, Italy\\
$^{3}$Department of Astronomy and Astrophysics, University of Toronto, 50 St. George Street, Toronto ON M5S 3H4, Canada\\
$^{4}$Department of Astronomy, Indiana University, Swain West, 727 E. 3rd Street, Bloomington, IN 47405, USA \\
$^{5}$European Southern Observatory, Karl-Schwarzschild-Strasse 2, 85748 Garching bei Munchen \\
$^{6}$Departamento de Ciencias Fısicas, Universidad Andres Bello, Republica 220, 837-0134 Santiago, Chile. 
}

\date{Accepted XXX. Received YYY; in original form ZZZ}

\pubyear{2020}

\begin{document}
\label{firstpage}
\pagerange{\pageref{firstpage}--\pageref{lastpage}}
\maketitle

\begin{abstract}
We exploit a combination of high-resolution {\it Hubble Space Telescope} and wide-field {\it ESO-VLT} observations to study the slope of the global mass function ($\alpha_G$) and its radial variation ($\alpha(r)$) in the two dense, massive and post core-collapse globular clusters M15 and M30. The available data-set samples the clusters' Main Sequence down to $\sim0.2 \ M_{\odot}$ {and the photometric completeness allows the study of the mass function between $0.40 \ M_{\odot}$ and $0.75 \ M_{\odot}$} from the central regions out to their tidal radii. We find that both clusters show a very similar variation in $\alpha(r)$ as a function of clustercentric distance. They both exhibit a very steep variation in $\alpha(r)$ in the central regions, which then attains almost constant values in the outskirts. Such a behavior can be interpreted as the result of long-term dynamical evolution of the systems driven by mass-segregation and mass-loss processes. We compare these results with a set of direct $N$-body simulations and find that they are only able to reproduce the observed values of $\alpha(r)$ and $\alpha_G$ at dynamical ages ($t/t_{rh}$) significantly larger than those derived from the observed properties of both clusters. {We investigate possible physical mechanisms responsible for such a discrepancy and {argue} that  both clusters might be born with a non-standard (flatter/{bottom-lighter}) initial mass function.}

%We investigate possible physical mechanisms responsible for such a discrepancy. Interestingly, our analysis suggests that both clusters might be born with a non-standard (flatter) initial mass function. This result has important implications not only for these two specific systems. In fact, assessing the universality and properties of the initial mass function is key for our understanding of stellar system formation and early evolution. 
\end{abstract}


%%%%%%%%%%%%%%%%%%%%%%%%%%%%%%%%%%%%%%%%%%%%%%%%%%

%%%%%%%%%%%%%%%%% BODY OF PAPER %%%%%%%%%%%%%%%%%%

\section{Introduction}

Globular clusters (GCs) are among the most populous, old, and dense stellar aggregates in the Universe and they play a crucial role in the study of many aspects of stellar evolution, stellar dynamics, and  the interplay between these two aspects  (see e.g. \citealt{heggie03}).

After an initial evolutionary phase likely driven by cluster environmental properties and stellar evolution, mainly related to high-mass star mass-loss and supernovae explosions \citep[see,e.g.][]{gieles06,kruijssen11,kruijssen12,renaud13,rieder13,mamikonyan17,li19}, the long-term dynamical evolution of a GC is driven by two-body relaxation and the external tidal field (see e.g. \citealp{heggie03} and references therein).   
%The main processes affecting a GC dynamical evolution include stellar evolution, two-body relaxation and the interaction with the external tidal field of the host galaxy. For GCs surviving the effects of the formation environment (see e.g.  Gieles et al. 2006; Kruijssen et al. 2011, 2012; Renaud & Gieles 2013; Rieder et al. 2013, Mamikonyan et al. 2017, Li \& Gnedin 2019, ***), stellar evolution is the initial driver behind cluster evolution and early expansion due to mass-loss from high-mass stars (see e.g. Chernoff \& Weinberg 1990, Fukushige \& Heggie 1995). After this early stages, evolution is driven by the effects of two-body relaxation and the external tidal field (see e.g. Heggie \& Hut 2003 and references therein).
{The effects of two-body relaxation drive more massive stars toward the cluster's center (mass segregation), while less massive stars migrate toward the cluster's outer regions.}
At the same time, this effect causes some stars to increase their energy and eventually escape the cluster. 

%The typical timescale in which two-body relaxation occurs scales with the number of stars \citep{spitzer87} and it is typically of the order of $1-2$ Gyr in GCs  \citep{meylan97}, which is significantly shorter than the average age of Galactic GCs ($\sim12$ Gyr), 
{The typical timescale  associated with the effects of two-body
relaxation is of the order of $1-2$ Gyr for most GCs \citep{meylan97}, which is significantly shorter than the average age of Galactic GCs ($\sim12$ Gyr),} thus suggesting that most of them have experienced quite a significant evolution.
The internal dynamics of stellar aggregates affect objects of any mass and its effects have been often probed by means of massive test stars, like blue straggler stars, binaries and millisecond pulsars \citep[e.g.][]{ferraro12,ferraro18,lanzoni07,lanzoni16,dalessandro09,dalessandro11,cadelano15,cadelano18,cadelano19}. 

The effects of mass segregation have also been traced by studying the radial variation of the slope of the stellar mass function \citep[MF;][]{beccari11,dalessandro15,webb17}. {In fact, the combined effects of mass segregation and star loss leads to the formation of gradients in the local (i.e. measured at different clustercentric distances) MF and to a gradual flattening of the global MF \citep[e.g.][]{vesperini97,baumgardt03,webb16}}. The effects of internal dynamics on variations in the local and the global MFs therefore need to be carefully considered in the interpretation of the observed differences between the MFs of various GCs. %and in any study aimed at understanding whether these differences can be explained entirely by the GCs dynamical history or require also primordial variations in the initial MF (IMF).}
Interestingly, by means of detailed comparison between observations and $N-$body models, we have shown \citep{webb16,webb17} that the combined measurements of the internal radial variation in the slope of the MF ($\delta_{\alpha}$) and its global value ($\alpha_G$) are able not only to trace the long-term dynamical evolution of a cluster, but also to put critical constraints on the system's initial MF (IMF). 
This constraint is of critical importance, as the IMF influences most of the observable properties (e.g. chemical composition, mass-to-light-ratio) of any stellar system, from star clusters to galaxies. Hence detecting variations in the IMF can provide deep insight into the processes by which stars form. While significant efforts have been made to study the IMF in a variety of different environments, no consensus has been reached regarding its universality \citep[e.g.][]{strader11,shanahan15}. 

As a part of a large program aimed at constraining the degree of dynamical evolution of GCs by analyzing their MF radial variations and studying possible variations of their IMFs \citep{dalessandro15,webb17}, here we present a detailed study of the MF of two dynamically evolved globular clusters: M15 (NGC7078) and M30 (NGC7099). Both clusters orbit the Galactic halo and have quite similar structural properties. 
They are both dense ($\log{\rho_c} (M_{\odot}/pc^3)  \sim7.5$ and $\sim5.9$ for M15 and M30, respectively) and relatively massive systems ($\sim10^5 \ M_{\odot}$; \citealp{baumgardt18}), hosting a stellar population with a very similar metallicity ($[Fe/H]\sim-2.3$; \citealp{carretta09,lovisi13}) and age ($\sim 13.25$ Gyr, \citealp{dotter10}). {Table~\ref{tab:prop} summarizes the main properties of the two systems.}
Based on the analysis of their blue straggler stars radial distribution \citep{ferraro12,ferraro18,lanzoni16,beccari19} both clusters appears to be dynamically very old. In addition,  studies of the density profiles \citep[][Beccari et al., in prep.]{noyola06,ferraro09} show that both clusters have already experienced core collapse. Undergoing core-collapse is another indication that these clusters are in advanced evolutionary stages and that their local and global MFs may have been significantly affected by evolutionary processes. Along the same line, 
in both clusters a double blue straggler star sequence has been observed \citep{ferraro09, beccari19}. Such a feature, which has been detected in several clusters now (namely M30, M15, NGC362 and possibly NGC1261; \citealp{ferraro09,dalessandro13,simunovic14,beccari19}) is interpreted as a clear indication of a quite advanced dynamical stage possibly connected with the core-collapse event.   
%They are therefore ideal systems to constrain models for the evolution of the MF and further explore the role of dynamics in the MF evolution.

The outline of the paper is the following: in Section~\ref{sec:obs} we present the data-set, the data reduction and artificial star test.  Section~\ref{sec:mf} reports on the MFs of the two clusters and their radial variation. In Section~\ref{sec:discuss} we compare the observational results with a set of $N$-body models. Finally, in Section~\ref{sec:conc} we draw our conclusions.

\begin{table}
\centering
 \caption{Main properties of the two clusters analyzed in this work. From top to bottom: mass, 2D half-mass and tidal radii, log of the central density, age, metallicity and log of the half-mass relaxation time.}
 \label{tab:prop}
 \begin{tabular}{c  c  c  c}
  \hline
  \hline
Param. & M15  &  M30 & Ref. \\
  \hline
$M \ (10^5 \ M_{\odot})$ & $4.99\pm0.05$  & $1.39\pm0.06$ & B19   \\ 
$r_{hm} \ (\arcsec)$ & $78\pm8$    & $92\pm9$ & B20,F09 \\
$r_t \ (\arcsec)$ & $750$    & $850$ & B20,F09 \\
$D$ ($Kpc$) & $10.22\pm0.13$ & $8.0\pm0.6$ & B19 \\
$\log \rho_c \ (M_{\odot} \ pc^{-3})$ & 7.5   & 5.9 & B19 \\
Age (Gyr) & $13.25\pm0.75$   & $13.25\pm0.75$ & D10  \\
$[Fe/H]$ & -2.3  & -2.3 & C09,L13   \\
$\log t_{rh}$ (yr) & $9.39\pm0.08$  & $9.11\pm0.09$ & This work \\
  \hline
  \hline
 \end{tabular}
\begin{tablenotes}
      \small
      \item References: B19 \citep{baumgardt19}; B20 (Beccari et al., in prep.); F09 \citep{ferraro09}; D10 \citep{dotter10}; C09 \citep{carretta09}; L13 \citep{lovisi13}.
    \end{tablenotes}
\end{table}

\section{Observations and data analysis} \label{sec:obs}

  \begin{figure*}
   \centering
   \includegraphics[scale=0.28]{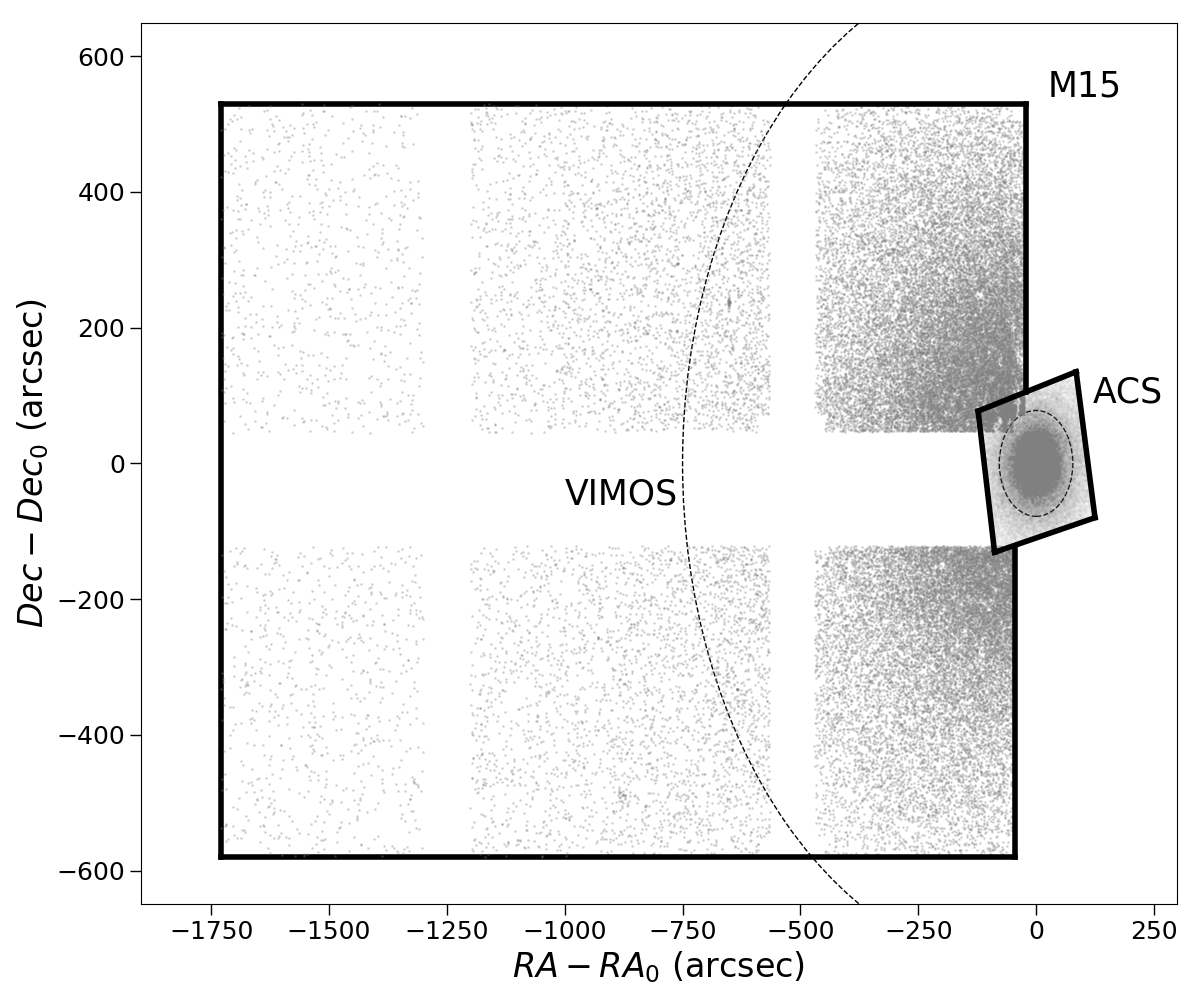}
   \includegraphics[scale=0.28]{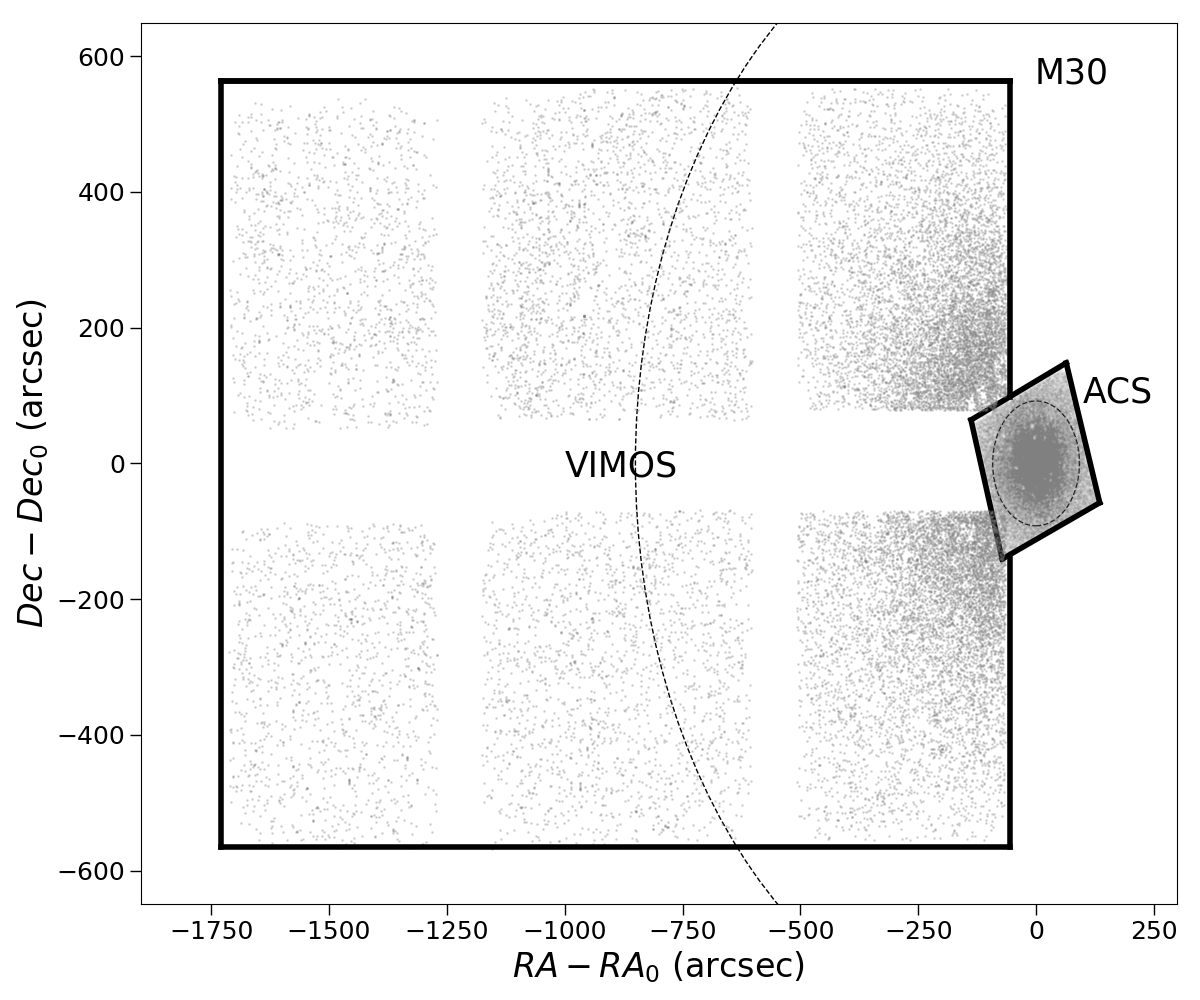}
   \caption{Field of views covered by the observations used in this work for M15 (left panel) and M30 (right panel). Each point represents a star. {White regions without stars correspond to the inter-chip gaps of the VIMOS detector. The inner and outer dashed circles are the cluster's projected half-mass and tidal radii, respectively (see Table~\ref{tab:prop}).}}
              \label{fig:fov}
    \end{figure*}
    
   \begin{figure*}
   \centering
   \includegraphics[scale=0.3]{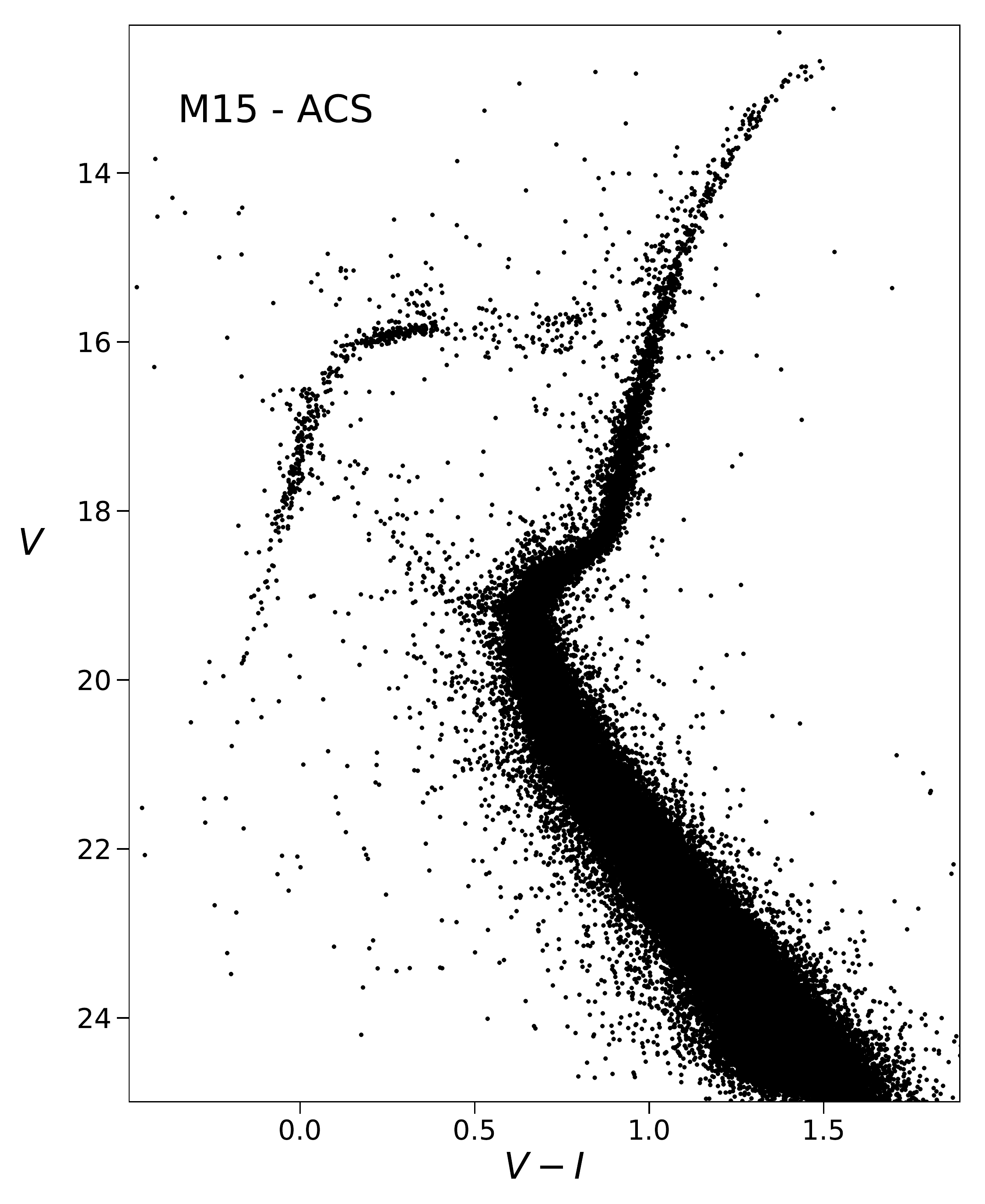}
   \includegraphics[scale=0.3]{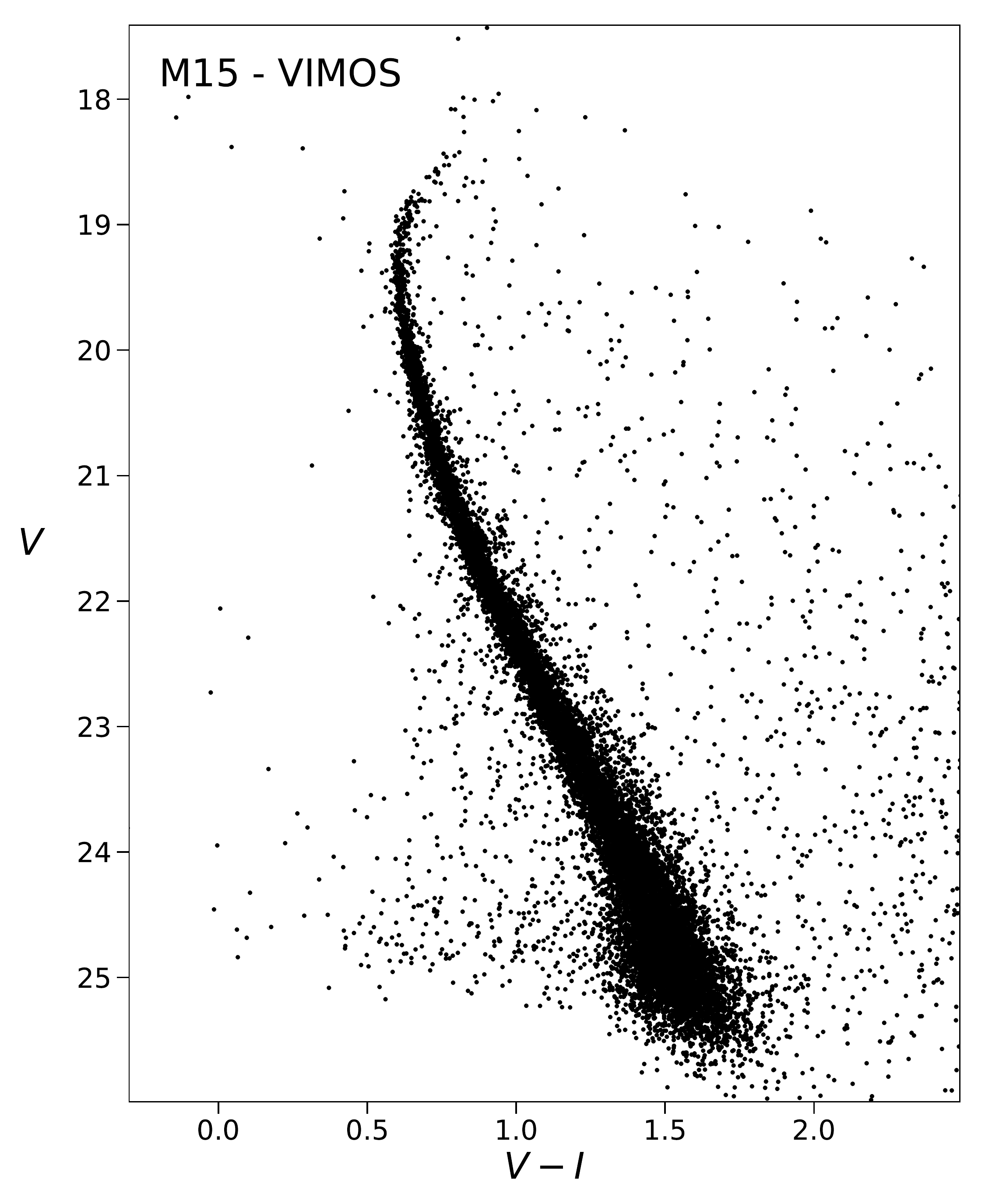}
   \caption{{\it Left panel:} $V$ vs ($V-I$) CMD of M15 as obtained from the high-resolution HST data-set by \citep{sarajedini07}. {\it Right panel:} $V$ vs ($V-I$) CMD of M15 as obtained from the ground based and wide field VIMOS data-set.}
             \label{fig:cmdM15}%
    \end{figure*}
    
   \begin{figure*}
   \centering
   \includegraphics[scale=0.3]{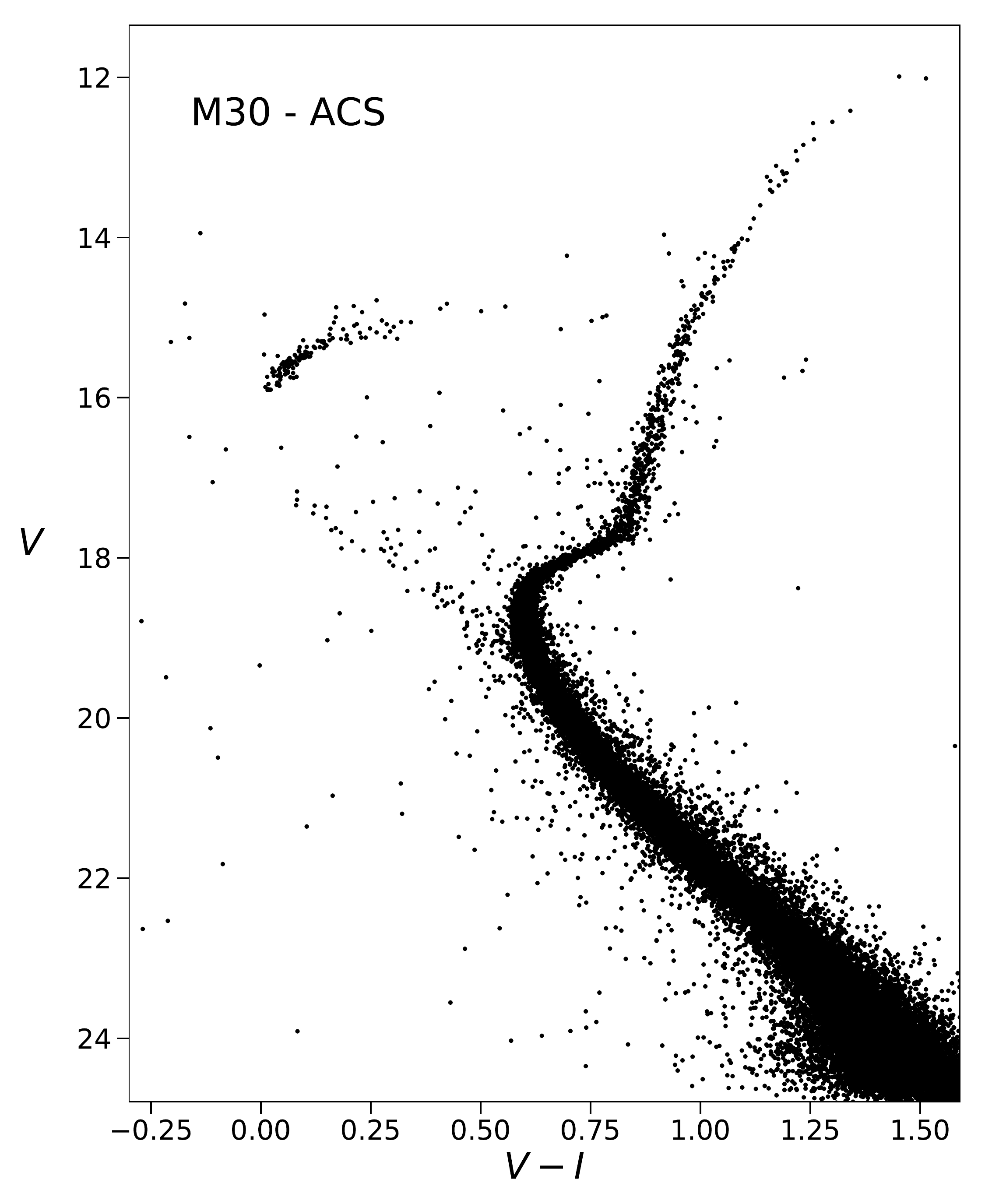}
   \includegraphics[scale=0.3]{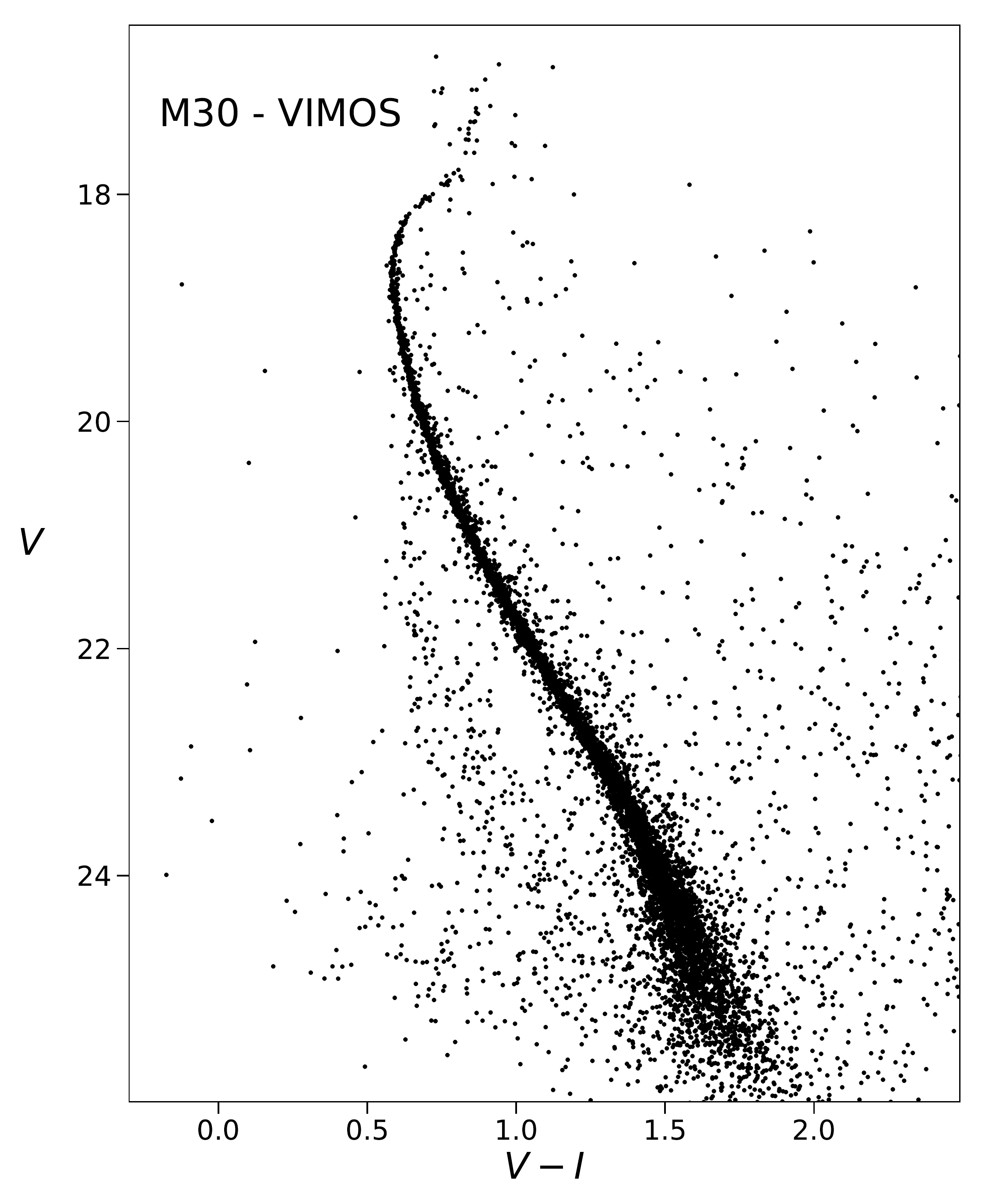}
   \caption{As in Figure~\ref{fig:cmdM15}, but for the M30 data-set.}
             \label{fig:cmdM30}%
    \end{figure*}
To study the radial variation of the MF along the entire cluster with adequate spatial resolution and photometric completeness, we combined high-resolution Hubble Space Telescope (HST) data with wide-field ground-based photometry.  
For M30 and M15 we made use of two twin data-sets and data-reduction strategies.

To sample the cluster's innermost and crowded regions we used the publicly available catalogs obtained as a part of the {\it ACS Treasury Survey of Galactic Globular Clusters} \citep{sarajedini07}. The survey was performed by using observations acquired with the Advanced Camera for Surveys aboard HST (proposal GO 10775; P.I.: Sarajedini). The data-set are composed of images equally split between the F606W and F814W bands and obtained with a combination of long and short exposure times (see \citealt{sarajedini07,anderson08} for details). The catalogs also provide calibrated Johnson V-band and I-band magnitudes, which we adopted throughout the whole work for homogeneity purposes with the wide-field catalogs. {These images approximately sample the cluster's extension till their half-mass radii (see Table~\ref{tab:prop}).}

The ground-based wide-field data-set samples each cluster's outer regions out to their tidal radii and consists of images acquired with the VIMOS camera mounted on the UT3 (Melipal) telescope at Paranal VLT/ESO observatory under Program ID: 097.D-0145(A) (PI: Dalessandro). 
In the case of M15, the data-set is composed of 12 images obtained with the Johnson V filter with exposure times of 305 s and 12 images obtained with the Johnson I filter with exposure times of 280 s. The images sample two overlapping fields of view (see Figure~\ref{fig:fov}), the first one centered at about $500\arcsec$ west from the cluster center and the second one at about $1250\arcsec$ west from the cluster. In the case of M30, the data-set is composed of 16 images per filter and we adopted the same combination of filters, exposure times and field of view coverage. The resulting total field of views extend beyond each cluster's tidal radii.
 
For each cluster, after correcting the images for bias and flat-field, we performed the photometric analysis independently on each image and on each chip of the detector by using {\rm DAOPHOT IV} \citep{stetson87}. As a first step, an adequate number of bright but not saturated stars have been chosen to model the point-spread function in each frame. This function was then applied to all the sources detected at $4\sigma$ above the background. We then created a master-list including all the sources detected in at least half of the images of each chip and, finally, a fit was forced in all the frames at the corresponding positions using {\rm DAOPHOT/ALLFRAME} \citep{stetson94}. For each star of the resulting catalog, we homogenized the magnitudes measured in different images and their weighted means and standard deviations have been adopted as the star's final magnitude and its related uncertainty.
The instrumental positions have been transformed to the absolute system by using the stars in common with the {\rm Gaia} Data Relese 2 archive \citep{gaia18}. The instrumental magnitudes have been reported to the Johnson photometric system by using the stars in common with the wide-field catalog described by \citet{stetson19} and \citet{ferraro09} for M15 and M30, respectively.

The total field of view covered by both the high resolution and wide field data-sets is shown in Figure~\ref{fig:fov}, while the obtained color-magnitude diagrams (CMDs) are plotted in Figure~\ref{fig:cmdM15} and Figure~\ref{fig:cmdM30}.

\subsection{Artificial star test}
\label{sec:completness}
To study the MF of the clusters and its radial variation it is necessary to take into account the completeness level of our catalogs for stars with different magnitudes and located at different distances from the cluster centers. 
To this aim, we run artificial star experiments. For the ACS data-set, we used the artificial star catalogs provided along with the main catalogs of the {\it ACS Survey of Galactic Globular Clusters} \citep[see Section~6 of][]{anderson08}. 

For the VIMOS data-set, we performed a large number of artificial star experiments following the prescriptions described in \citealt{dalessandro15} (see also \citealp{bellazzini02}). We created a list of artificial stars with a V-band input magnitude extracted from a luminosity function modeled to reproduce the observed ones in the same filters and extrapolated beyond the limiting magnitude. Then, to each of these stars, we assigned an I-band magnitude by interpolating along the mean ridge line of the clusters. These artificial stars were added to the real images by using the {\rm DAOPHOT/ADDSTAR} software. The photometric reduction process and the point-spread function models used for the artificial star experiments are exactly the same as described in Section~\ref{sec:obs}. This process was iterated multiple times and, in order to avoid ``artificial crowding'', stars were placed into the frames in a regular grid composed of $38\times38$ pixel cells (corresponding approximately to ten times the typical FWHM of the point spread function) in which only one artificial star for each run was allowed to lie.
At the end of the runs, about 100000 and 150000 were simulated for the entire field of view covered by the M15 and M30 VIMOS data-set, respectively. 

\begin{figure*}
   \centering
   \includegraphics[scale=0.3]{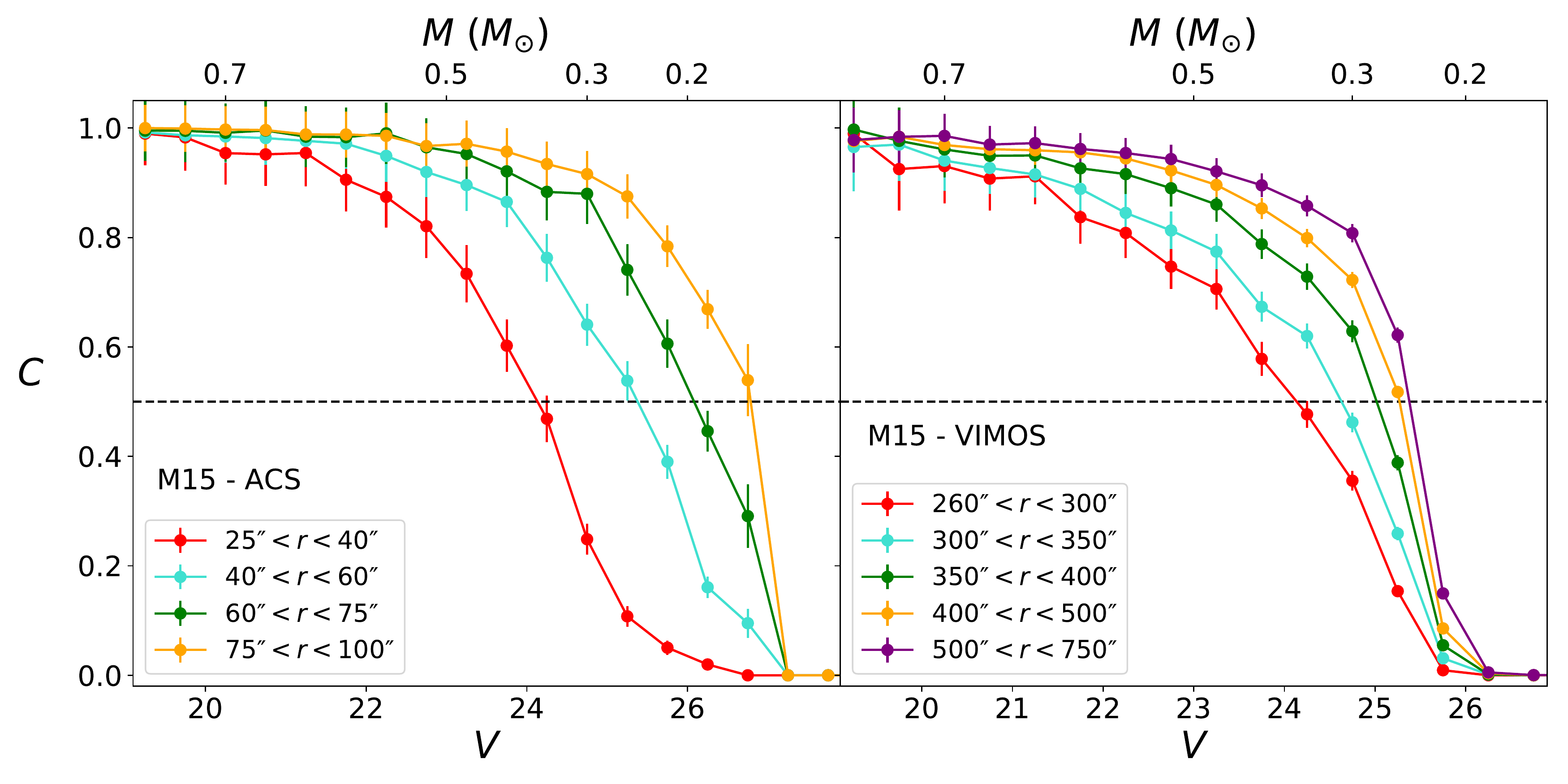}
   \includegraphics[scale=0.3]{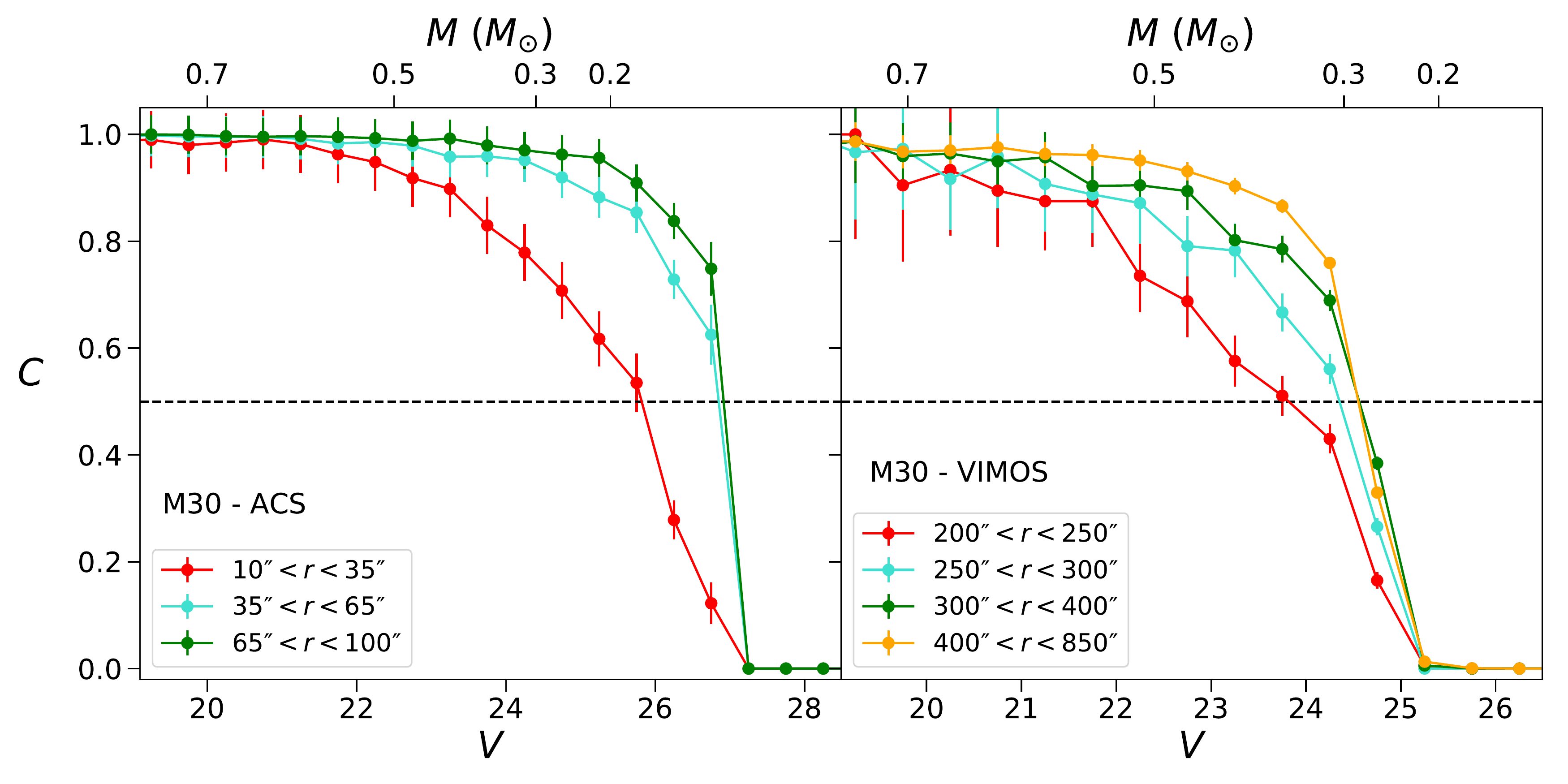}
   \caption{Completeness curves as a function of the V-band magnitudes (stellar mass) for both M15 (top panels) and M30 (bottom panels) and separately shown for the ACS data-set (left panels) and the VIMOS data-set (right panels). Different curves, extracted at different radial distances from the cluster center, are plotted. {The dashed horizontal lines mark the lowest completeness level ($C=0.5$) considered in the data analysis.}}
              \label{fig:compl}%
    \end{figure*}

 {A completeness value $C=N_o/N_i$, defined as the ratio between the number of stars recovered at the end of the artificial star test ($N_o$) and that of stars actually simulated ($N_i$), was assigned to each star by using the following approach. To account for the effect of crowding (and therefore of the distance of the stars from the cluster center) on the completeness, for each star the completeness $C$ was derived by using only objects located within a radial bin centered on the location of the star and with a width of $5\arcsec$ and $50\arcsec$ for the ACS and VIMOS data-sets, respectively. The bin widths were chosen as a compromise between having enough statistics and sampling a limited radial extension. Since the completeness level strongly depends on the stellar magnitude, we evaluated $C$ considering only simulated objects within a 0.5 large magnitude bin, centered on the V-band magnitude of each star. Finally, the uncertainties $\sigma_C$ on the completeness value of each star were computed by propagating the Poissonian errors.  Figure~\ref{fig:compl} shows the variation of $C$ as a function of the V-band magnitude in a selection of radial bins.}

%For these selected stars, we then evaluated $C$ as a function of the star magnitudes.  In such a way, completeness curves, as those shown for example in Figure~\ref{fig:compl}, are produced and the completeness level of each star of a given magnitude and at a given distance from the cluster center is recovered by interpolating its completeness curve. 

\section{Mass Function}
\label{sec:mf}
  \begin{figure*}
   \centering
   \includegraphics[scale=0.3]{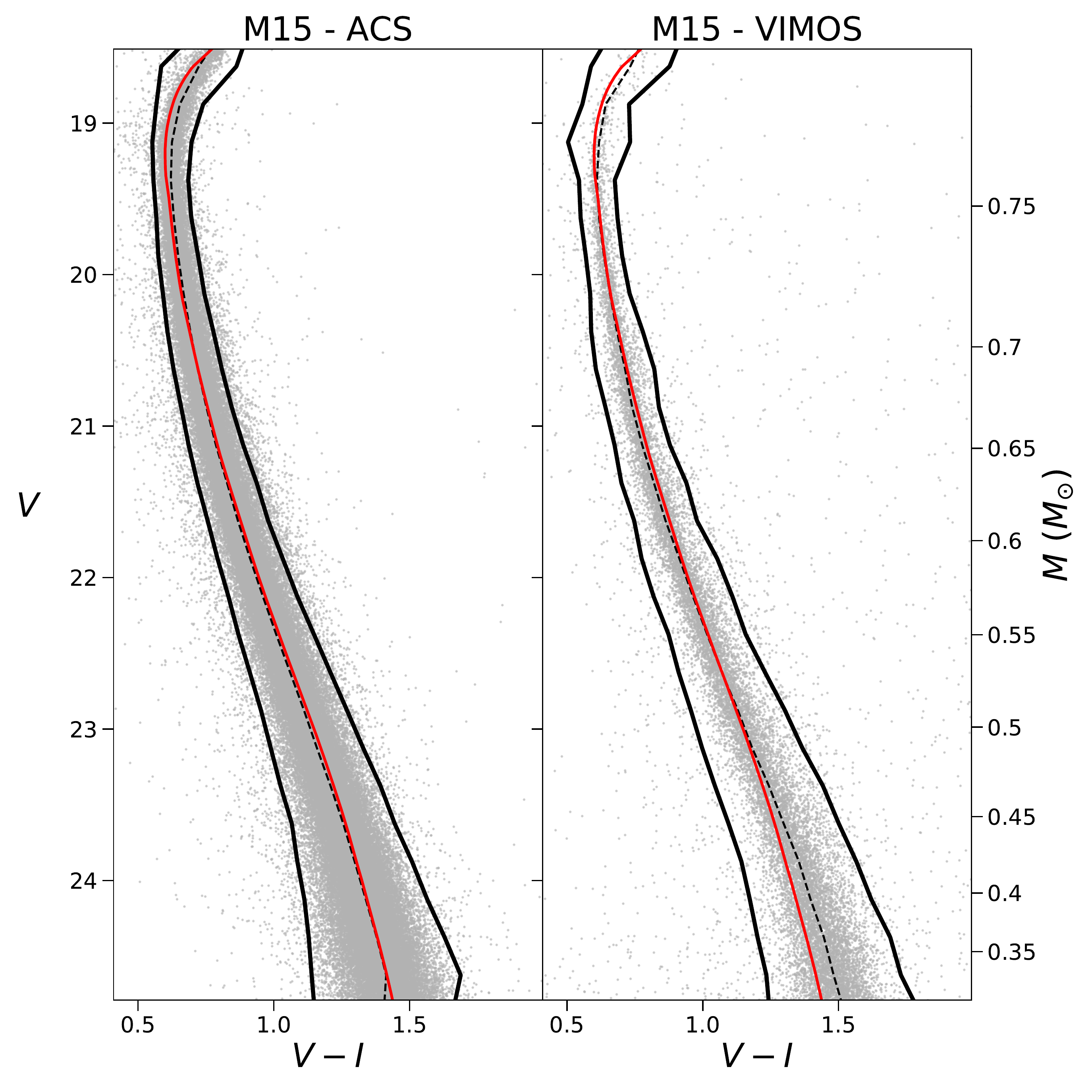}
    \includegraphics[scale=0.3]{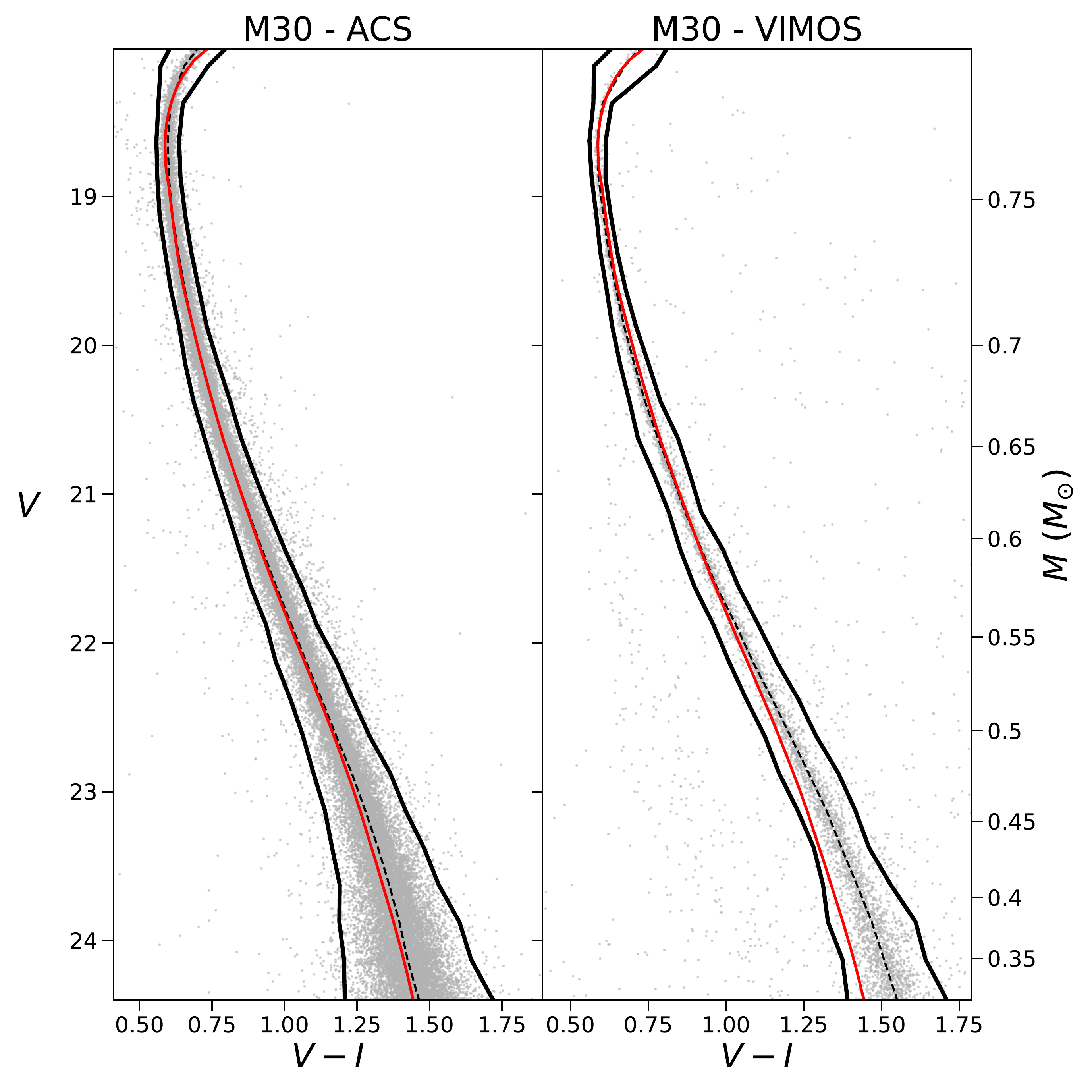}
   \caption{{\it Left panel:} CMDs obtained from the ACS and VIMOS data-set of M15. Black curves enclose the selected bona-fide main sequence cluster stars for which we derived the mass. The dashed black curve is the cluster mean-ridge line. The red curve is the adopted isochrone model from which we derive stellar masses at different V-band magnitudes. {\it Right panel:} same as in the left panel, but for the case of M30.}
              \label{fig:massa}%
    \end{figure*}
To derive the cluster MF, we first selected as bona-fide cluster members those stars lying along the observed and well-defined main sequence of the two clusters.  To this end, we built the cluster mean ridge lines by computing the $3\sigma$-clipped average color of stars within different bins in the magnitude range $18.5<V<26$ and $18<V<26$ for M15 and M30, respectively. In both cases, we adopted a 0.25 mag bin width and in each bin we selected as bona-fide cluster stars those located within $3\sigma$ the measured average color (see black curve in  Figure~\ref{fig:massa}). 
%Then, we converted the stellar magnitudes of the selected stars into masses by projection onto an appropriate isochrone model. 
We used isochrones from the {\rm Dartmouth Stellar Evolution Database} \citep{dotter07} for a stellar population with an age of 13.25 Gyr \citep{dotter10} and with a metallicity of $[Fe/H]=-2.3$ and $[\alpha/Fe]=+0.2$, suitable for both clusters \citep{carretta09,lovisi13}. Absolute magnitudes were converted to the observed frame by adopting a distance modulus of $(m-M)_0=15.17$ and a color excess of $E(B-V)=0.08$ in the case of M15, while we adopted a distance modulus of $(m-M)_0=14.72$ and a color excess of $E(B-V)=0.05$ in the case of M30 \citep{ferraro99}. Figure~\ref{fig:massa} shows that the isochrones nicely reproduce the observed CMDs, {although a small deviation is visible in the low-luminosity regions of both the clusters' main sequence. This, however, has a negligible effect in the following analysis}. We applied an interpolation to derive the masses as a function of the V-band magnitude, {as predicted by the isochrone models}. As can be seen, both data-sets cover a broad range of masses {from $0.76 \ M_{\odot}$ (turn-off mass) down to $\sim 0.3-0.2 \ M_{\odot}$}.

  \begin{figure*}
   \centering
  \includegraphics[scale=0.3]{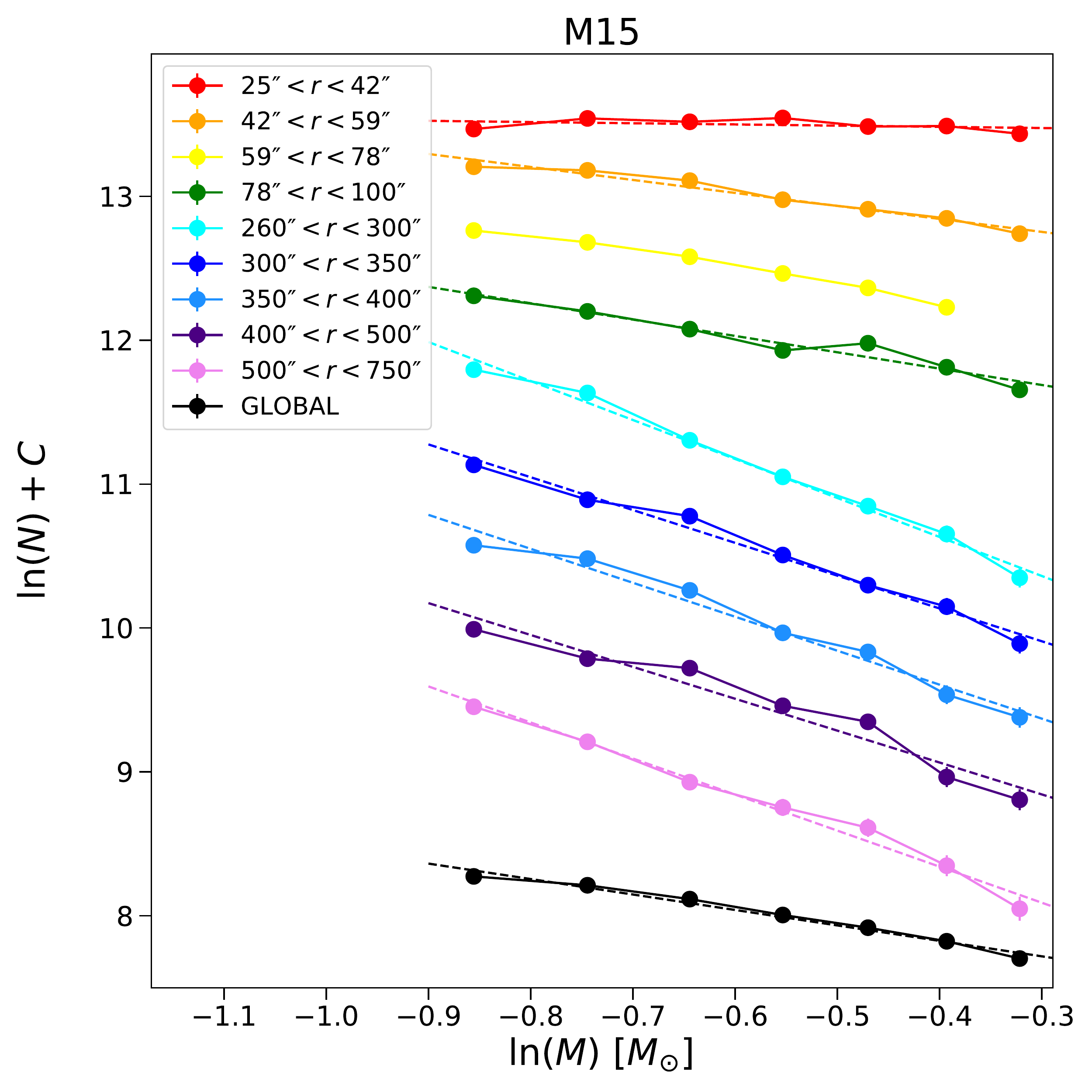}
   \includegraphics[scale=0.3]{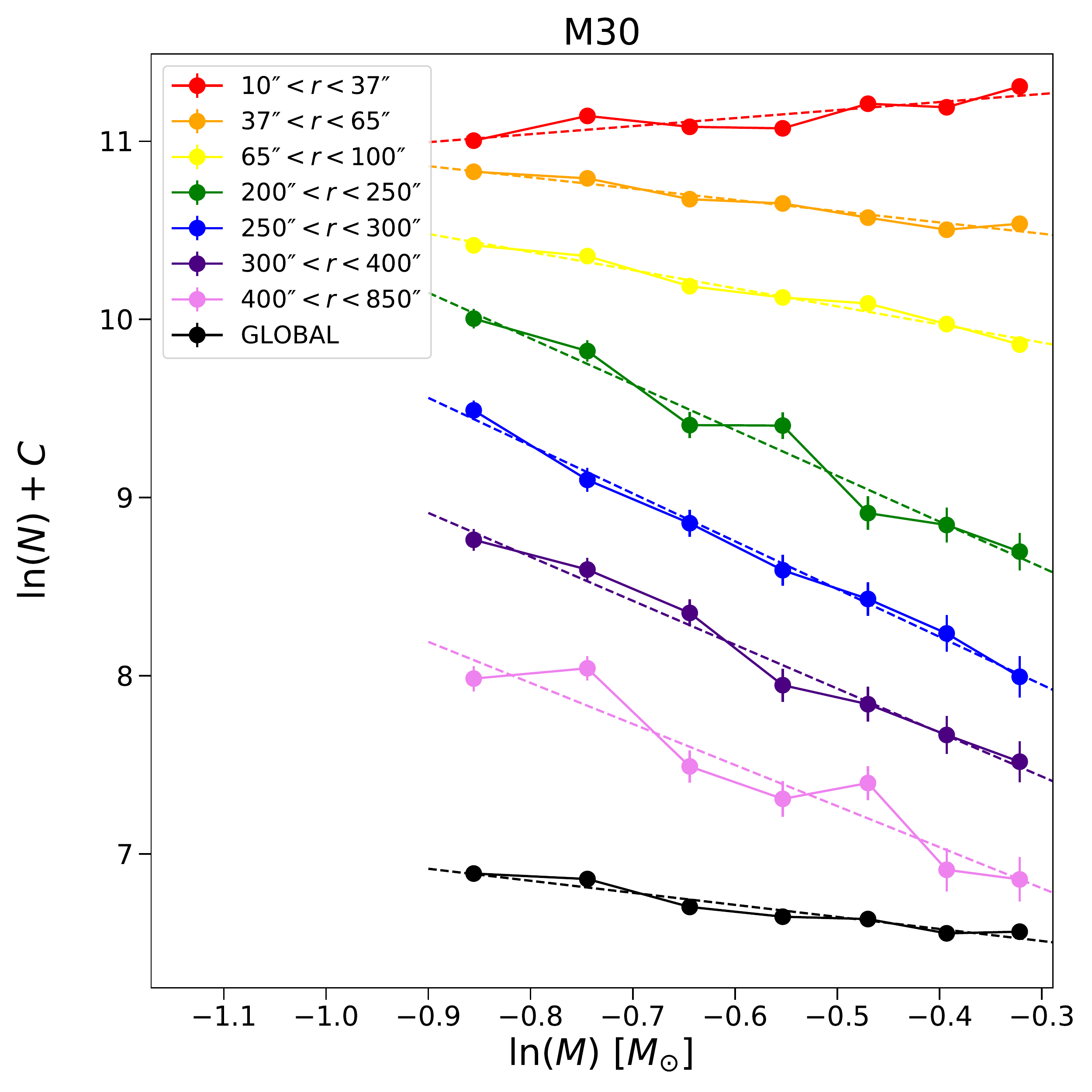}
   \caption{{\it Left Panel:} Stellar MFs obtained from the M15 data-set used in this work. Different colors correspond to different radial bins, as specified in the legend. An arbitrary constant was added to the different MFs for clarity. The lines represent the linear best fit obtained for each of the MFs. {\it Right Panel:} same but for the case of M30.}
              \label{fig:massfunc}%
    \end{figure*}

 To compute the stellar MFs of each cluster, we counted the number of stars located at different distance bins from the cluster center.  
In order to maximize the reliability of the results, we restricted the analysis only to the stars with completeness larger than $50\%$. Thus in the case of M15 we considered only stars in the mass range $0.40 M_{\odot}$-$0.75 M_{\odot}$ and located only at distances larger than $25\arcsec$ and $250\arcsec$ from the cluster center of the ACS and VIMOS data-set, respectively. In the case of M30, we considered stars in the same mass range but located at distances larger than $10\arcsec$ and $200\arcsec$ from the cluster center of the two data-sets. {The completeness corrected number of stars and its uncertain in each radial and mass bin are $N_{corr}=\sum_{i}^{N_{obs}} C_{i}^{-1}$ and $\sigma_{N_{corr}}=\sqrt{\sum_{i}^{N_{obs}} (\sigma_{C_i} / C_{i})^2}$, respectively, where $N_{obs}$ is the number of stars observed in a given bin, $C_i$ the completeness of the $i^{th}$ star and $\sigma_{C_i}$ its uncertainty derived as described in Section~\ref{sec:completness}}.
The MFs evaluated at different radial bins are plotted in Figure~\ref{fig:massfunc}. The number and widths of the radial bins have been set to sample approximately an equal number of stars, which is 24000 and 8500  for the ACS data-set and 2000 and 900 for the VIMOS data-set of M15 and M30, respectively.  To quantify the contamination by field interlopers, we evaluated, in different mass bins, the stellar density in an outer radial bin located a distances larger than $900\arcsec$ and $1300\arcsec$ from the centers of M15 and M30, respectively, beyond the cluster tidal radii (see Table~\ref{tab:prop}). {Then, in each radial and mass bin we subtracted to $N_{corr}$ the completeness corrected number of interlopers expected on the basis} of the bin area and of the measured stellar density in the outer bin. 
%We finally applied the Poisson statistic to quantify the uncertainties in the number of stars counted in each radial and mass bin. 
Results of this analysis are shown in Figure~\ref{fig:massfunc}. As expected, the MF slopes decrease significantly moving from the cluster centers to the outskirts. 

\begin{figure*}
   \centering
   \includegraphics[scale=0.35]{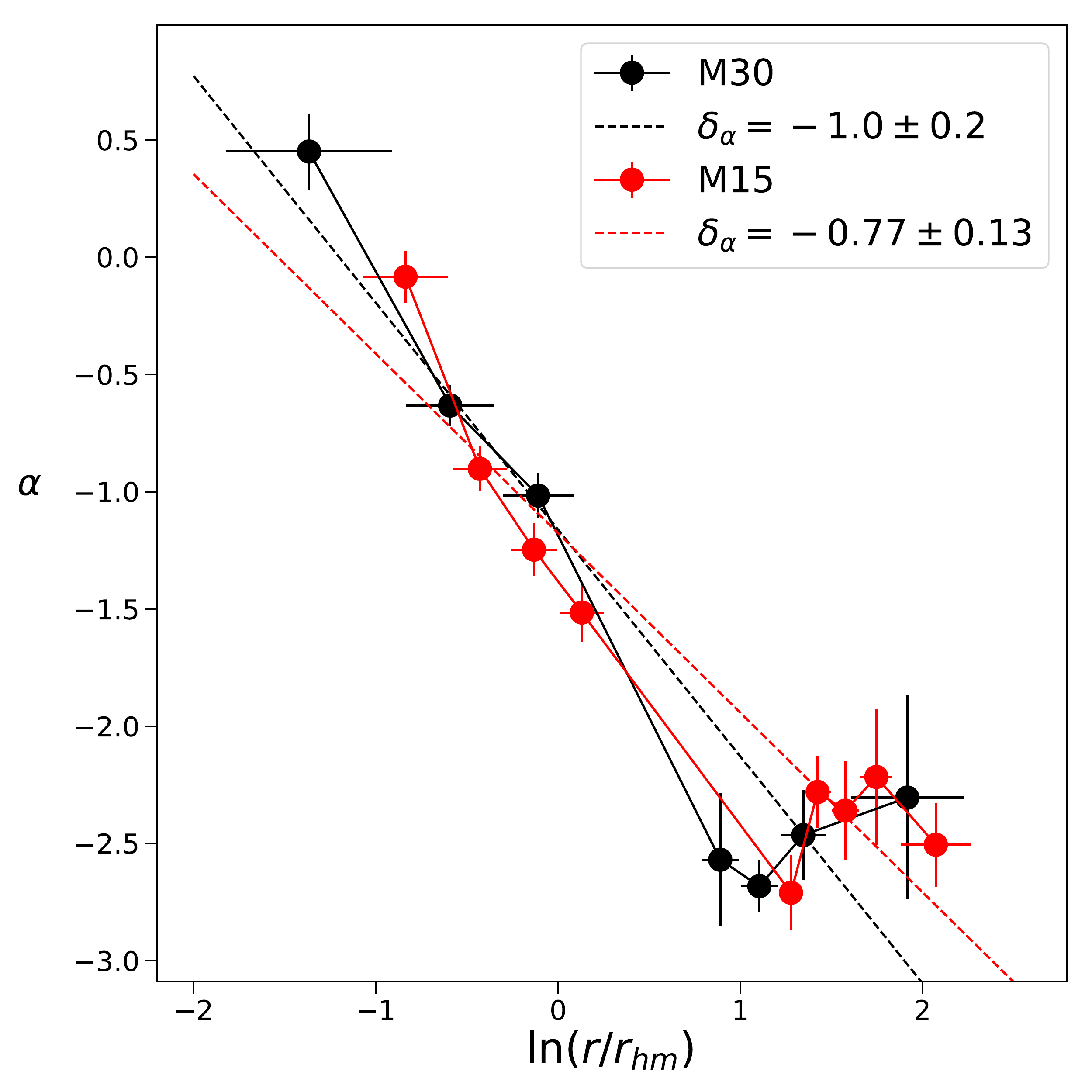}
   \caption{Variation of the slope of the MF with respect to the logarithmic distance from the cluster center expressed in units of half-light radii. The dashed lines are the best linear fit to the observed slope variations and their slopes $\delta_\alpha$ are reported in the legend. Black and red points and lines are for M30 and M15, respectively. }
              \label{fig:alpha}%
    \end{figure*}

To quantify the radial variation in the slope of the stellar MFs, we performed a linear fit to each of the measured stellar MFs reported in Figure~\ref{fig:massfunc}. The resulting slopes are reported in Table~\ref{tab:alpha} and they are plotted as a function of the logarithmic distance from the cluster center expressed in units of half-mass radii $r_{hm}$ in Figure~\ref{fig:alpha}. For the 2D projected half-mass radius, $r_{hm}$, we adopted the values quoted in Table~\ref{tab:prop}.  
Both clusters show quite similar radial variations of their slopes suggesting a similar dynamical evolution. Indeed, the central regions covered by the HST data-set are characterized by a rapidly steepening of the slopes for increasing clustercentric distances. 
Such a trend is the expected outcome of the mass segregation process. On
the other hand, the cluster outskirts, mapped through the VIMOS
data-set, are characterized by nearly constant slopes that can be
explained as the combined effect of mass segregation and preferential
loss of low-mass stars in the external region of both clusters due to the interaction with the Galaxy potential. 

\subsection{Main sources of uncertanties}
\label{sec:errors}
{In the following we discuss three potential sources of uncertainties and their impact on the derived MFs. These are: the uncertainties on the photometric completeness assigned to each star, the accuracy on the assignation of stellar masses along the main-sequence and, finally, the role of binaries.
\begin{itemize}
\item {To assess the impact of the photometric completeness uncertainties in the derivation of the MF slopes, we repeated several times, for each radial bin, the derivation of the MFs as described above. During each iteration, the completeness of each star was randomly drawn from a normal distribution centered on its completeness level and with a standard deviation equal to its uncertainty. At the end of the procedure, we obtained the MF slope distributions and we computed their $16^{th}$, $50^{th}$ and $84^{th}$ percentiles to quantify the spread introduced by the completeness uncertainties. Such a spread turned out to be as large as $\sim0.005$, thus negligible with respect to the uncertainties quoted in Table~\ref{tab:alpha} and due to the residuals of the linear fit of the MFs.}

\item{The results here obtained are based on the stellar masses derived through the mass-luminosity relation predicted by the \citet{dotter07} isochrones. Different models differing in terms of various assumption about stellar evolution, underlying chemical mixture and bolometric corrections could lead to slightly different mass-luminosity relation. To quantify the effect this may have on the derived MF slopes, we repeated the whole analysis deriving the stellar masses using isochrones generated from the Victoria-Regina Isochrone Database \citep{vandenberg2014}, the BaSTI stellar evolution models \citep{pietrinferni04,
  pietrinferni06} and the PARSEC database \citep{marigo17}. For each of these databases, we extracted an isochrone with the same stellar age and metallicity used before.
Results for both clusters show basically no differences in the radial variation of the MF slopes ($\delta_\alpha$, see Section~\ref{sec:discuss}). Also, the global MF slopes $\alpha_G$ obtained using the Victoria-Regina isochrone are basically the same as derived with the Dartmouth Stellar Evolution model. 
On the contrary, the $\alpha_G$ values obtained by using the BaSTI and PARSEC models turned out to give systematically flatter MFs (up to $\delta \alpha \sim 0.5$) than those reported in Table~\ref{tab:alpha}.}
%we obtained $\alpha_{G,BaSTI}=-0.85\pm0.09$ and $\alpha_{G,PARSEC}=-0.43\pm0.14$ in the case of M15, while we obtained $\alpha_{G,BaSTI}=-0.36\pm0.17$ and $\alpha_{G,PARSEC}=0.03\pm0.19$ in the case of M30. These values further increase the discrepancy between the N-body models and the observations. 
%Therefore, the derivation of masses through different isochrones (thus different mass-luminosity relation) do not change our general results. However, it is important to stress that great caution should be used when comparing MF slopes derived using different isochrone models.

\item {Finally, to evaluate the impact of binaries, we repeated the analysis selecting only the stars in the blue side of the mean ridge lines shown in Figure~\ref{fig:massa}. This region is in fact expected to be populated almost exclusively by single stars. The general results are unchanged and deviations from the values quoted in Table~\ref{tab:alpha} are $\lesssim15\%$. Therefore binary systems do not have a significant impact on our analysis. Indeed, both cluster's host a small binary fraction \citep[$\sim2-3\%$;][]{milone12} that is likely centrally segregated due to the advanced stage of dynamical evolution of both the systems.}
\end{itemize}
}

\begin{table}
\centering
 \caption{Slopes of the stellar MFs derived at different distances $r$ from the cluster centers.}
 \label{tab:alpha}
 \begin{tabular}{c  c | c  c }
  \hline
  \multicolumn{2}{c}{M15} \vline &  \multicolumn{2}{c}{M30}\\
  \hline
  $r$(\arcsec) & $\alpha$  &  $r$(\arcsec) & $\alpha$\\
  \hline
  25-42 & $-0.08\pm0.11$ & 10-37 & $0.45\pm0.16$\\
  42-59 & $-0.90\pm0.10$ & 37-65 & $-0.63\pm0.09$  \\
  59-78 & $-1.25\pm0.11$ & 65-100 & $-1.02\pm0.10$ \\
  78-100 & $-1.52\pm0.13$ & 200-250 & $-2.6\pm0.3$ \\
  260-300 & $-2.71 \pm0.16$ & 250-300 & $-2.7\pm0.1$\\
  300-350 & $-2.28 \pm0.15$ & 300-400 & $-2.5\pm0.2$ \\
  350-400 & $-2.4 \pm 0.2$ & 400-850 & $-2.3\pm0.4$\\
  400-500 & $-2.2 \pm 0.3$ & GLOBAL & $-0.68\pm0.10$\\
  500-750 & $-2.51 \pm 0.18$ \\
  GLOBAL & $-1.07\pm0.08$ \\
  \hline
 \end{tabular}
\end{table}

\section{Comparison to $N$-body Simulations}
\label{sec:discuss}

\subsection{The $N$-body simulation set}
For a more quantitative interpretation of the observational results, we will compare them to $N$-body simulations from \citet{webb16} that model the evolution of star clusters in a Milky Way-like external tidal field. The simulations were performed using the direct $N$-body code NBODY6 \citep{aarseth03}, with each star cluster's initial conditions generated assuming a Plummer density profiles \citep{plummer11} out to a cutoff of ten half-mass radii. In order to consider both initially compact and extended clusters, we will be comparing the observations to model clusters with initial half-mass radii $r_{hm,i}$ of 1.2 and 6 pc. 

Both model clusters initially consists of $10^5$ stars, with masses drawn from a \citet{kroupa93} IMF in the range 0.1 - 50 $M_{\odot}$ . Hence their initial masses are approximately $6 \times 10^4 M_{\odot}$. Stars then evolve with time according to the stellar evolution algorithms of \citet{hurley00} assuming a metallicity $Z=0.001$.

The Milky Way-like potential within which both clusters are evolved is made up of a point-mass bulge, a \citet{miyamoto75} disk, and logarithmic halo. The bulge has a mass of $1.5 \times 10^{10} M_{\odot}$ while the disk has a mass of $5 \times 10^{10} M_{\odot}$ and scale radii of $a = 4.5$ kpc and $b = 0.5$ kpc. The logarithmic halo is scaled such that all three components combine to yield a circular velocity of 220 km/s at 8.5 kpc. Both model clusters have circular orbits at 6 kpc from the center. %, which are comparable to M15 slightly eccentric orbit at a Galactocentric distances $\sim 4-10$ kpc, while are quite oversimplified for the eccentric orbit of M30 at distances $\sim 1.5-8$ kpc from the Galaxy center (BAUMGARDT19).

%The observed radial variation of the MF slopes as a function of the clustercentric distance shown in Figure~\ref{fig:alpha} are indicative of a very advanced dynamical evolution in these systems. To derive more quantitative insights on the dynamical state of M30 and M15, we compared the observed results to $N$-body simulations. 

\subsection{Comparing models and observations}
\label{sec:comparing}
In measuring the radial variation of the simulated cluster's MFs, we project the positions of each stars onto a random two-dimensional plane and only include stars in the same mass range and fields of view as our observed data-set. {For each observed cluster, we have determined the boundaries of the fields of view in terms of the cluster’s half-mass radius.  These boundaries are then used to determine what subset of stars in each N-body simulation should be considered when measuring the MF and its radial variation, with the half-mass radius of the cluster at the current time-step being used to scale the boundaries.} Following \citet{webb17} we present the cluster evolution in terms of the linear slope of the radial variation of the MF slopes, defined as: $\delta_\alpha=d \alpha(r)/d \ln(r/r_{hm})$, which is a good measure of a GC's degree of mass-segregation, and the slope of the global MF $\alpha_G$, which is a proxy of the mass lost by a cluster \citep{vesperini97,webb15}, with respect to the ratio between the cluster stellar age and its current half-mass relaxation time $t/t_{rh}$. 

First of all, we measured $\delta_\alpha$ and $\alpha_G$ for both clusters. $\alpha_G$ was measured by counting all the stars in a single radial bin covering the entire radial extension considered for the local $\alpha$ measurements (see Table~\ref{tab:alpha} and Figure~\ref{fig:alpha}), thus in regions where the completeness is always larger than 50\%. {Please note that, due to the field of view geometry (see Figure~\ref{fig:fov}), the outer radial bins cover smaller radial extensions than the inner ones and by consequence the latter have a larger weight in the derived $\alpha_G$ values. Therefore extra-caution should be used when comparing the values here derived with those obtained using different data-sets.}. We found $\delta_\alpha=-0.77\pm0.13$ and $\alpha_G=-1.07\pm0.08$ for the case of M15, while we found $\delta_\alpha=-1.0\pm0.2$ and $\alpha_G=-0.68\pm0.10$ for the case of M30. 
{While this is the first time that $\delta_\alpha$ is measured for these two clusters, we can compare the $\alpha_G$ values here derived with those quoted in previous works. \citet{paust10} found $\alpha_G=-0.92\pm0.06$ for M30, while no values is reported for M15. \citet{sollima17} found instead $\alpha_G=-1.16\pm0.06$ and $\alpha_G=-0.72\pm0.02$ for M15 and M30, respectively, {while \citet{ebrahimi20} report $\alpha_G=-1.00\pm0.04$ and $\alpha_G=-0.80\pm0.03$ for M15 and M30, respectively}. Finally, the compilation of \citet{baumgardt18} quotes $\alpha_G=-0.53$ and $\alpha_G=-1.02$ for M15 and M30, respectively. Therefore, there is a general reasonable agreement between our and previous works. However, we stress that all these literature values were obtained through a combination of observations and modelling, and, for the values reported in \citet{baumgardt18}, also considering stars in a mass range slightly different than that adopted in this work. On the other hand, our results are based exclusively on observations, {although the outer regions are not uniformly sampled and thus the $\alpha_G$ values are likely biased toward the inner regions of the cluster}.
}

To evaluate the ratio $t/t_{rh}$, we adopted $t=13.25\pm0.75$ Gyr for both the clusters \citep{dotter10}, while we derived $t_{rh}$ following \citet{spitzer71}:
$$
t_{rh}= 2.054\times 10^6 \ yr \ \frac{M^{\frac{1}{2}}}{<m>} \frac{r_{hm}^{\frac{3}{2}}}{\ln(0.4\frac{M}{<m>})}
$$
where $M$ is the cluster’s mass {in units of solar masses}, $r_{hm}$ is the projected half-mass radius (see Table~\ref{tab:prop}) {in pc units} and $<m>$ is the mean stellar mass (assumed to be, {as in the \citealp{harris10} catalog, $\frac{1}{3}M_{\odot}$}). %{Here we used the projected 2D half-mass radius instead of the 3D one in order to compute the relaxation time in the same exact way as done in the models.}
We found $t_{rh}=2.5\pm0.5$~Gyr for M15 and $t_{rh}=1.3\pm0.3$~Gyr for M30, thus implying $t/t_{rh}=5.3\pm1.1$ and $t/t_{rh}=10\pm2$ for M15 and M30, respectively.

\subsection{M15}

\begin{figure*}
   \centering
   \includegraphics[scale=0.52]{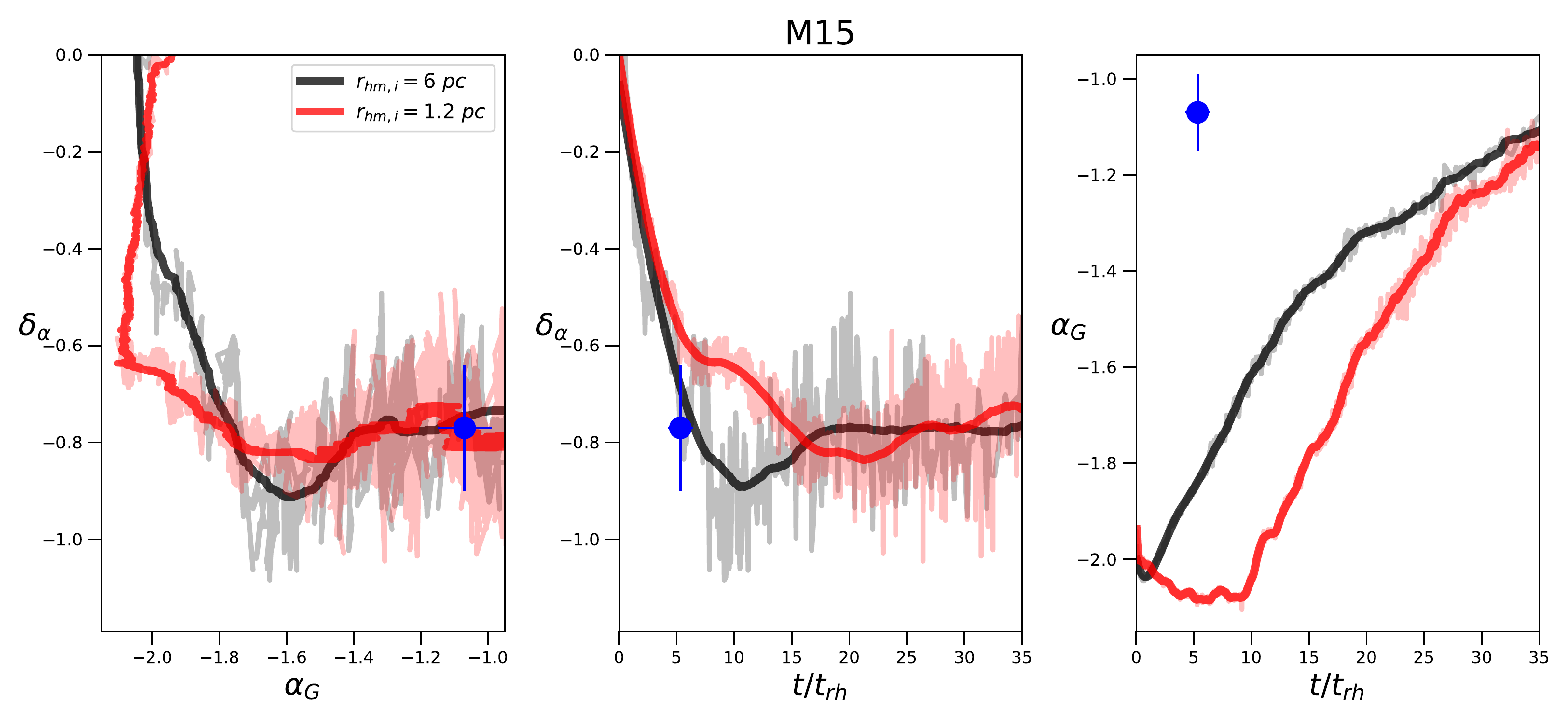}
   \caption{The left panel shows the evolution of slope of the best linear fit to the observed variation in the slope of the stellar MF ($\delta_\alpha$) with respect to the slope of the global MF ($\alpha_G$). The middle and right panels show instead the evolution of $\delta_\alpha$ and $\alpha_G$ with respect to the ratio between the cluster age and the instantaneous half-mass relaxation time ($t/t_{rh}$).  The red and black lines correspond to the smoothed evolution of direct $N$-body star cluster simulations with initial half-mass radii of 1.2 pc and 6 pc. The shaded areas show instead the real values of the simulations. For comparison purposes, the blue points mark the positions of M15 in this parameter space. }
              \label{fig:fitM15}%
    \end{figure*}

In the left panel of Figure~\ref{fig:fitM15}, we show the model cluster evolution in the $(\delta_\alpha,\alpha_G)$ plane, together with the measured positions of M15 in this parameter space.
A nice match is reached with both the extended and compact cluster simulations. However, we note that M15 falls in a region of this diagram in which the expected evolution of $\delta_{\alpha}$ is largely insensitive to $\alpha_G$ variations. {This is essentially due to the fact that $\delta_\alpha$ stops decreasing since segregation in the core has stopped and tidal stripping in the outer regions prevents $\alpha$ from decreasing further. However, the global $\alpha$ will continue to increase as low-mass stars escape the cluster.} For this reason we also compared the behavior of both $\delta_{\alpha}$ and $\alpha_G$ with respect to $t/t_{rh}$. The middle and right panels of Figure~\ref{fig:fitM15} show that while the observed value of $\delta_{\alpha}$ is well reproduced by the models at the estimated $t/t_{rh}$, $\alpha_G$ is significantly flatter than predicted. Indeed, at the cluster corresponding $t/t_{rh}$, the models predict an $\alpha_G$ value around -1.6 for the extended cluster and around -2 for the compact one. The model is able to match the observed $\alpha_G$ value only at significantly later stages of the evolution, around $t/t_{rh}\sim30$,  which is a factor of 3 larger than what estimated for M15. Any reasonable uncertainties on both age and relaxation time would be hardly able to account for such a large difference.
{It is also important to note that had we used the $\alpha_G$ values derived adopting other isochrones (see Section~\ref{sec:errors}), the discrepancy between observations and simulations would have been even more severe.}

{The flatter $\alpha_G$ found in our observational data would suggest that M15 
has lost significantly more mass than predicted by our simulations.
However, the strength of the external tidal field adopted in the
simulations is similar to that inferred from the orbit of M15 (which is
currently on a slightly eccentric orbit at a Galactocentric distances
around $4-10$ kpc; \citealp{baumgardt19}). In addition, as shown in \citet{webb16},
 the dependence of the evolution of both $\delta_\alpha$ and $\alpha_G$ on
the orbital properties is not sufficient to explain the observed
discrepancy. Exceptional events, such as tidal shocks or interactions with molecular clouds, that could increase the mass-loss rate over a relatively short period of time, are not taken into account by the models. However, these are rare events that are unlikely to explain the observed discrepancy. {In any case, further simulations are needed to firmly confirm the effects of such events in the evolution of both the radial and global MF slopes. As far as the possible effects of primordial mass segregations are concerned, \citet{webb16} have shown that for old GCs and the mass range considered in this study a broad range of different degrees of primordial mass segregation do not have a strong effect on the value of $\alpha_G$ after one Hubble time. 
%The presence of primordial mass-segregation could in principle lead to a faster evolution of $\alpha_G$ over time since initially segregated clusters undergo a significant early expansion due to mass-loss of high-mass stars. During this phase, a large number of low-mass stars escape the cluster and thus $\alpha_G$ rapidly increase. However, as pointed out by \citet{webb16}, primordial mass-segregation have a negligible effect on the evolution of $\delta_\alpha$ and $\alpha_G$ of relatively high-mass stars (e.g. $0.3<M/M_{\odot}<0.8$) as those considered in the MFs presented in this work. Furthermore, the presence of primordial mass-segregation could also introduce a discrepancy between the predicted and observed values of $\delta_\alpha$ over $t/t_{rh}$ and $\alpha_G$, although such a discrepancy is expected to be larger for dynamically young clusters.} 
Finally, our two simulations cover a broad range of initial half-mass radii 
%in the degree of the tidal filling
%factor ($r_{hm}/r_J=0.03$ for $r_{hm}=1.2$ pc and $r_{hm}/r_J=0.14$ for $r_{hm}=6$ pc -- where $r_J$ is the Jacobi radius}) 
and given the
variation of the evolution of $\delta_\alpha$ and $\alpha_G$ with this
parameter, it is unlikely that the observed discrepancy could be
explained by a different choice for the initial half-mass radius.} %{In fact, a more compact and denser cluster at birth would results in a slower evolution of $\alpha_G$, thus further increasing the observed discrepancy. On the other hand, a faster evolution of $\alpha_G$ could be obtained with a very expanded and low-density cluster at birth, which however would not be able to reach the present-day degree of mass-segregation observed in M15. }

A different IMF, flatter than the \citet{kroupa93} IMF adopted here could
easily remove the difference between observations and models, as this
cluster would start the evolution with a larger value of $\alpha_G$. {\it Our
analysis thus suggests that a different IMF might be the most likely
explanation to the observed ($\alpha_G$, $\delta_\alpha$, $t/t_{rh}$) trends.}}

\subsection{M30}
\begin{figure*}
 \centering
 \includegraphics[scale=0.52]{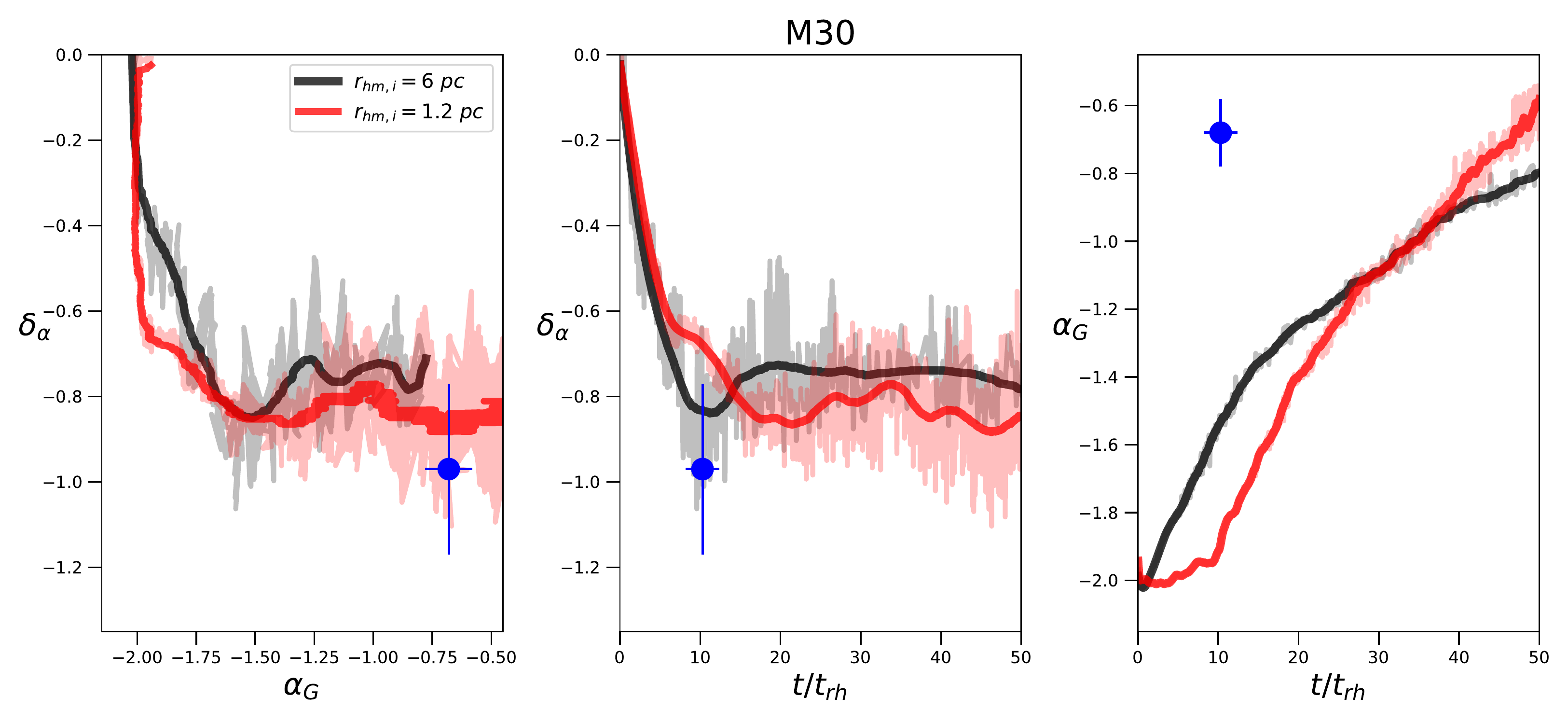}
 \caption{As in Figure~\ref{fig:fitM15}, but for the case of M30.}
 \label{fig:fitM30}%
 \end{figure*}
Figure~\ref{fig:fitM30} shows the same diagnostic plots we used for M15 but for the case of M30. Also in this case the derived values of ($\delta_{\alpha}$, $\alpha_G$) are nicely reproduced by the simulations. % while a marginally good match is obtained instead in the ($\delta_{\alpha}$, $t/t_{rh}$) frame.  Indeed,  the $\delta_\alpha$ observed value is slightly steeper than predicted by the model. This could suggests that M30 formed even more compact than our simulated cluster\footnote{Indeed, \citet{webb16} already ruled out the possibility of binaries, primordial mass segregation, black holes, or a different IMF causing a unique evolution of $\delta_{\alpha}$.}.  
However, as in the case of M15, the simulations are not able to match the observations in the ($\alpha_G$, $t/t_{rh}$) diagram where actually the discrepancy is even larger than for M15.  The $\alpha_G$ value measured for M30 is reached by the simulations at a very late stages of the evolution, around $t/t_{rh}\sim40$, significantly larger than the value of this ratio determined from observational data. {Also in this case, the derivation of $\alpha_G$ assuming different stellar evolution models (see Section~\ref{sec:errors}) would further increase the mismatch.}
%The same arguments invoked to explain the mismatch in M15 can be followed here too with some extra caution. In fact, M30 more likely suffered a higher degree of mass-loss than the simulated cluster. In fact, M30 has a very eccentric orbit in a distance range $\sim 1.5-8$ kpc from the Galaxy center, thus with a  pericenter smaller than the distance adopted in our simulations. However again, different assumptions on the cluster orbit are not expected to have a significant impact on the evolution of $\delta_\alpha$ and $\alpha_G$ \citep{webb16}. A more extended cluster at birth could again at least partially help to solve the mismatch, but in this case would produce a more significant discrepancy between models and observations in the ($\delta_{\alpha}$, $\alpha_G$) and ($\delta_{\alpha}$, $t/t_{rh}$) diagrams.Therefore also in this case, the comparison between observations and simulations would suggest that M30 was born with non-standard and flatter IMF. 

{The same arguments discussed for M15 also apply to M30. One difference
to note is that M30 has a very eccentric orbit in a distance range $\sim1.5-8$ kpc from the Galaxy center, thus with a pericenter smaller than the distance adopted in our simulations. However, again, the dependence
on the cluster's orbit (see \citealp{webb16}) is unlikely to
account for the differences between observational data and numerical
models revealed by our analysis. {In any case, given the differences between the real and simulated orbits, the discrepancy could be partially due to a larger degree of mass-loss due to the cluster highly eccentric orbit. It also worth mentioning that, on the basis of its current orbit, \citet{massari19} suggested that M30 formed in the Gaia-Encaladus dwarf galaxy. This would mean that M30 could have experienced a milder tidal field than in the Galaxy, during the early stage of its evolution. However, this should not have a significant impact in our analysis, since the merger event with the dwarf Gaia-Enceladus dates back to $\sim10$ Gyr ago \citep{kruijssen20} and therefore the cluster spent most of its life in the Milky Way.} {\it Given all this, we suggest that also for this cluster a different, flatter/bottom-lighter, IMF is likely to be required to reconcile
observational data with theoretical models.}}

\section{Conclusions and Discussion}
\label{sec:conc}

We used a combination of deep, high-resolution and wide-field optical observations of the dynamically old Galactic GCs M15 and M30 to investigate their dynamical evolution in terms of the radial variation of their stellar MF along the whole cluster radial extensions. Both clusters reveal a quite similar variation of the MF slopes with respect to the clustercentric distance. In fact, the inner regions (approximately within $r_{hm}$) show a progressive steepening of the MF slopes while moving away from the cluster centers. On the other hand, the outer regions (approximately from $5 r_{hm}$ to their tidal radii) are characterized by almost constant MF slopes. This trend is the expected outcome of the long-term dynamical evolution driven by two-body encounters and progressive mass-loss due to the cluster interactions with the Galaxy.

We compared the observed results with a set of direct $N$-body models, following the cluster evolution in a Milky Way-like potential, assuming a standard \citet{kroupa93} IMF. Such a comparison has been performed by means of two powerful indicators of the cluster degree of mass-segregation and mass-loss: the radial variation of the MF slope ($\delta_\alpha$) and the slope of the global MF ($\alpha_G$), respectively. We found that the models are able to nicely reproduce the measured values in the ($\delta_\alpha,\alpha_G$) diagram. However, in both M15 and M30, the dynamical state of the cluster as traced by the ($\delta_\alpha$, $t/t_{rh}$) and ($\alpha_G$, $t/t_{rh}$) is reproduced only at significantly later stages of the evolution, when the ratio between the cluster age to the instantaneous half-mass relaxation time is $\sim3-4$ times larger than the measured ratios for the two clusters. As largely discussed in \citet{webb16}, different assumptions about the initial binary fractions and dark remnants (and their retention), as well as on the cluster's orbit cannot account for such a differences. {On the other hand, also the uncertainties on the observed quantities cannot explain the discrepancy between observations and simulations.}. 
The results obtained in this paper would suggest that the most likely explanation to such a significant discrepancy is the adoption of a non universal IMF, flatter/bottom-lighter than the one assumed by models. 

{A correlation between the global MF slopes and the half-mass relaxation time (and the ratio of the age to the half-mass relaxation time) of a sample of Galactic GCs, with dynamically older clusters showing flatter MF slopes, was recently found by \citet{sollima17} and \citet{ebrahimi20}. Such a correlation may be difficult to reconcile with significant variations of the IMF and it has been argued it is likely to result from the effects the dynamical evolution alone.
However, the situation appears to be more complicated and the presence of IMF variations cannot be excluded. In general, the star forming environment should play an important role in shaping the IMF of stellar systems \citep[see e.g.][for some theoretical and observational studies about this topic]{silk77,strader11,giersz11,henault20,ebrahimi20,kroupa20}. {In this respect, it is important to point out that other studies have noted that the discrepancy between theoretical predictions and observations of metal-rich GC mass-to-light ratios might be due to a non-standard IMF, either bottom-light (i.e. fewer low-mass stars) or top-light MFs (i.e. fewer dark remnants)
(see \citealt{strader11} and \citealt{henault20}, in which other possibilities in alternative to a non-universal IMF are also discussed).
The results obtained in the this work would suggest possible IMF variation also at very metal-poor regime.}}

%However, the results obtained here suggest a more complicated scenario and the presence of IMF variations cannot be excluded. In this respect, it is important to point out that the star forming environment should play an important role in shaping the IMF of stellar systems and, in general, metal-rich clusters are expected to have flatter IMFs than metal-poor clusters \citep[see][for both theoretical and observational studies about this topic]{silk77,strader11,giersz11,henault20,ebrahimi20}. Interestingly, while we found hints of an IMF flatter than a \citet{kroupa93} for two very metal-poor clusters, the coherent application of this method to a sample of GCs differing in terms of metallicity will open to the possibility to confirm the existence of such a correlation and thus to better quantify the role of the environment in star formation.}

{One aspect that has not been investigated yet, neither theoretically nor
observationally, concerns the possible variations in the IMF of multiple
stellar populations observed in almost all Galactic GCs. Different observations 
suggest that the chemically anomalous second population (i.e. Na-rich, O-poor) of stars form in a
compact system more segregated with respect to the first population of stars (see \citealp{lardo11,dalessandro19}) as predicted by multiple population formation scenarios (see \citealp{bastian18,gratton19} for recent reviews). The
implications of the different formation environments of stars in the
first and second populations are still unknown and the connection with
the possible evidence of a non universal IMF will require further
studies.
}

More in general, constraining the IMF of stellar clusters have key implications on our understanding of their formation process and early evolution, with strong impact on the early enrichment undergone by stellar clusters, gas consumption efficiency, stellar cluster initial mass and their contribution to building-up of the Galactic halo.

\section*{Acknowledgements}
{The authors thank the anonymous referee for the careful reading of the paper and the constructive comments.
We also thank B. Lanzoni for useful discussion}. MC and ED acknowledge financial support by the project Light-on-Dark granted by MIUR through PRIN2017-000000 contract.

Based on observations collected at the European Southern Observatory under ESO programme 097.D-0145(A).

Based on observations made with the NASA/ESA Hubble Space Telescope, obtained from the Data Archive at the Space Telescope Science Institute, which is operated by the Association of Universities for Research in Astronomy, Inc., under NASA contract NAS 5$-$26555. These observations are associated with program GO~10755.

The data underlying this article will be shared on reasonable request to the corresponding author.

%%%%%%%%%%%%%%%%%%%%%%%%%%%%%%%%%%%%%%%%%%%%%%%%%%



%\bibliographystyle{mnras}


%%%%%%%%%%%%%%%%%%%%%%%%%%%%%%%%%%%%%%%%%%%%%%%%%%

%%%%%%%%%%%%%%%%% APPENDICES %%%%%%%%%%%%%%%%%%%%%

%\appendix


%If you want to present additional material which would interrupt the flow of the main paper, it can be placed in an Appendix which appears after the list of references.

%%%%%%%%%%%%%%%%%%%%%%%%%%%%%%%%%%%%%%%%%%%%%%%%%%

% Don't change these lines
\bsp	% typesetting comment
\label{lastpage}
\end{document}